\documentclass[12pt]{article}
\title{Point Form Electrodynamics and the Gupta-Bleuler
Formalism}
\author{W. H. Klink\\Department of Physics and Astronomy\\University of
Iowa, Iowa City, Iowa, USA}
\begin{document}
\maketitle
\begin{abstract}
The Gupta-Bleuler formalism for photons is generalized
by choosing the representation of the little group for
massless particles, the two dimensional Euclidian group, to be
the four dimensional nonunitary representation obtained by restricting
elements of the Lorentz group to the Euclidian group.  Though the little
group representation is nonunitary, it is shown that the representation
of the Poincar\'{e} group is unitary.  Under Lorentz transformations
photon creation and annihilation operators transform as irreducible
representations of massless particles, and not as four-vectors.  As a
consequence the polarization
vector, which connects the four-vector potential with creation and
annihilation operators, is given in terms of boosts, coset
representatives of the Lorentz group with respect to the Euclidian
group.  Several polarization vectors (boost choices) are worked out,
including a front form polarization vector.  The different boost choices
are shown to be related by the analogue of Melosh rotations, namely
Euclidian group transformations\footnote{PACS, 14.70.B}.
\end{abstract}
\section{Introduction}
The goal of this series of papers is to construct a relativistic
many-body theory of hadrons using the point form of relativistic quantum
mechanics, in which all interactions are vertex interactions,
arising from products of fields evaluated at the space-time point zero. 
In the first of this series of papers \cite {a} such vertex interactions
were constructed for the hadronic part of the mass operator.  The second
paper showed how to construct one-body current operators for arbitrary
spin particles \cite{b}.  In order that the electromagnetic interaction
also be a vertex interaction, the hadronic currents should be coupled to
the four-vector potential operator in such a way that the two contracted
operators give a scalar density under Lorentz transformations.  Moreover
the four-vector potential operator should transform as a four-vector
under Lorentz transformations, yet be constructed out of photon creation
and annihilation operators that transform as one-photon states under
Lorentz transformations.

To construct such a vertex, it will be necessary to generalize the
Gupta-Bleuler formulation \cite{c} for photons.  For unlike the usual
Gupta-Bleuler formulation, the photon creation and annihilation
operators should themselves transform under the appropriate irreducible
representation of the Poincar\'{e} group, namely the massless
representations for which the little group is $E(2)$, the two dimensional
Euclidean group.  The reason is that in order to have an electromagnetic
interaction that is a vertex interaction, the polarization
vector should be a boost, connecting the creation and annihilation
operators with the four-vector potential operator, as is the case for
massive particles with spin (see the previous paper, reference \cite{b},
Eq.99).

The problem here is well-known;  since the two dimensional Euclidean
group is noncompact, it has only one dimensional or infinite dimensional
unitary representations.  The one dimensional representations are usually
thought of as providing the relevant representations for a massless spin
one particle like the photon and the two polarization states of the
photon arise as a consequence of parity.  Then it is simple to construct
photon creation and annihilation operators with two helicities that
transform in the same way as one-photon states.  However a problem arises
when one wishes to construct a four-vector potential field from these
photon creation and annihilation operators, for the "four-vector"
potential will not transform as a four-vector under Lorentz
transformations \cite{d}.

The solution to this problem is also well-known, and goes under the
heading of the Gupta-Bleuler formalism \cite{c};  one introduces photon
creation and annihilation operators with four components, and eliminates
the two spurious components using gauge invariance.  However, if the four
components transform as a four-vector under Lorentz transformations,
there is no natural way to link the polarization vector to boosts, as is
done in constructing fields for massive particles.  This paper will show
how photon creation and annihilation operators transforming under $E(2)$
representations naturally link to the four-vector potential field in such
a way that the polarization vector is given as a boost, coming from the
nonunitary four dimensional representation of the Lorentz group.

One of the advantages of such a
procedure is that any boost (coset representative of  the Lorentz group
with respect to E(2)) can be used as a polarization vector.  As will be
shown, the usual choice of polarization vector corresponds to a helicity
boost.  But other choices, such as a front form boost discussed in the
appendix, can also be used. In section 2, motivated by induced
representation theory, the relevant one-photon states and wave functions
are obtained, and the analogue of Wigner rotations for massless particles
is derived.  Though the four dimensional representation of the Euclidian
group is nonunitary, it will be shown that the full Poincar\'{e} group
representation is unitary.  A many-photon theory is generated by photon
creation and annihilation operators which have the same Poincar\'{e}
transformation properties as the single particle photon states. In
section 3 the four-vector potential operator is defined in terms of the
photon creation and annihilation operators, the link being the
polarization vector, that is, a boost Lorentz transformation.  What is
important here is that the four-vector potential transform as a
four-vector under Lorentz transformations, so that, when contracted with
the current operators of the previous paper, the electromagnetic vertex
be a Lorentz scalar.  The section closes with a discussion of gauge
transformations for photon creation and annihilation operators and free
field four-vector operators. 
\section{Photons and the Gupta-Bleuler Formalism}
As is well known the little group for all massless particles is the
Euclidian group in two dimensions.  A simple proof using $SL(2,C)$ is
given in the appendix.  If $\Lambda$ is an element of $SO(1,3)$, the
proper Lorentz group, then the two dimensional Euclidian group, $E(2)$,
can be defined as the subgroup of the proper Lorentz group leaving a
standard four vector invariant:
\begin{eqnarray}
E(2)&=&\{\Lambda \in SO(1,3)|\Lambda k^{st}=k^{st}\},
\end{eqnarray}
where $k^{st}:= (1,0,0,1)$ is the standard four vector.

To get a Poincar\'{e} group representation for massless particles, it is
necessary to choose a representation for the little group.  Wigner
\cite{e} (and others, for example Weinberg, reference \cite{d}, page 71)
choose the degenerate one dimensional unitary representation for
$E(2)$ in which the $E(2)$ translations are trivial, and the rotation
angle $\phi$ is represented by $e^{i{\lambda}\phi}$, with $\lambda$ equal
to plus or minus one.  Parity connects the plus or minus one helicities,
so that photon states can be written as $|k, \lambda=\pm 1>$, corresponding
to plus or minus helicity states.  The problems with this construction
have  to do with gauge invariance and the link to the four-vector
potential operator, which has four components which do not transform among
themselves under Lorentz transformations (see reference \cite{d}, page
250).

To get around these problems, in this paper another
representation for the little group $E(2)$ will be chosen, namely the
representation given by the group itself as defined in Eq.1.  In
terms of the
$SL(2,C)$ definition of the two dimensional Euclidian group given in the
appendix, Eq.1 can be thought of as a four dimensional nonunitary
representation of $E(2)$, for which the representation of an $e_2$ element
is  written as
$\Lambda(e_2)$, to indicate the Lorentz transformation representing the
Euclidean group element $e_2$.  The elements of the Euclidian group in this
representation can be written explicitly as
\begin{eqnarray}
\Lambda(\phi)&=&\left[\begin{array}{cccc}1&0&0&0\\
0&\cos\phi&-\sin\phi&0\\0&\sin\phi&\cos\phi&0\\0&0&0&1\\
\end{array}\right]\\
\Lambda(\vec{a})&=&\left[\begin{array}{cccc}1+|\vec{a}|^2/2&a_x&a_y&-|\vec{a}|^2/2\\
a_x&1&0&-a_x\\-a_y&0&1&a_y\\|\vec{a}|^2/2&a_x&-a_y&1-|\vec{a}|^2/2\\
\end{array}\right],\
\end{eqnarray}
where $\vec{a}$ gives the two translations of the Euclidean group.

Any Lorentz transformation can be written as a boost times a Euclidian
group element, where the boost $B(k)$ is a Lorentz transformation, that is
a coset representative of $SO(1,3)$ with respect to $E(2)$.  Boosts have
the property of sending $k^{st}$ to the four vector $k$:  $k=B(k)k^{st}$,
from which it follows that $k{\cdot}k:=k^T g k=k^\alpha k_\alpha=0$.  $g$
is the Lorentz metric matrix, $g:=diag(1,-1,-1,-1)$.  The usual boost
choice for massless particles is the helicity boost,
$B_H(k)$, which, as will be shown in section 3, gives the usual
polarization vector:
\begin{eqnarray}
B_H(k)&=&R(\hat{k}) \Lambda_z(|\vec{k}|),\\
&=&\left[\begin{array}{cccc}ch\chi&0&0&sh\chi\\
sh\chi\hat{k}&\hat{k}_1&\hat{k}_2&ch\chi\hat{k}\\
\end{array}\right]\
\end{eqnarray}
where $R(\hat{k})$ is the rotation matrix taking $\hat{z}$ to the
unit vector $\hat{k}$, $\Lambda_z(|\vec{k}|)$ is a Lorentz
transformation about the $z$ axis with
 $|\vec k|=e^\chi$ and \\
 $\hat{k}_1=(cos\phi cos\theta,sin\phi cos\theta,
-sin\theta)$, $\hat{k}_2=(-sin\phi,cos\phi,0)$.  Another boost choice, a
front form boost, is given in the appendix, Eq.42.

Since any Lorentz transformation can be written as $\Lambda= B(k)
\Lambda(e_2)$, it follows that the product of two Lorentz
transformations, namely $\Lambda B(k)$ can again be decomposed into such
a product, namely $B(k^{'}) \Lambda(e_W)$:
\begin{eqnarray}
\Lambda B(k)&=&B(\Lambda k) \Lambda(e_W)\\
\Lambda(e_W)&=&B^{-1}(\Lambda k)\Lambda B(k),\
\end{eqnarray}
where $k^{'}=\Lambda k$ is found by applying Eq.6 to $k^{st}$.
$\Lambda(e_W)$ is the massless analogue of a Wigner rotation(defined for
representations of the Poincar\'{e} group when particles have nonzero mass). 
For a given boost, such as a helicity boost defined in Eq.4, the
massless Wigner transformation is defined in Eq.7.

With these tools it is possible to define photon states with four
degrees of polarization and investigate their transformation properties
under Lorentz and space-time transformations:
\begin{eqnarray}
U_{e_2}|k^{st} \alpha>&=&\sum|k^{st}
\alpha^{'}>\Lambda_{\alpha^{'}\alpha}(e_2),\\
|k,\alpha>:&=&U_{B(k)}|k^{st},\alpha>,\\
U_\Lambda|k,\alpha>&=&U_\Lambda U_{B(k)}|k^{st},\alpha>\nonumber\\
&=&\sum|\Lambda k,\alpha^{'}>\Lambda_{\alpha^{'}\alpha}(e_W),\\
U_a|k,\alpha>&=&e^{ik.a}|k,\alpha>,
\end{eqnarray}
where $\Lambda(e_W)$ is the Wigner transformation defined in Eq.7;  in
particular if the Lorentz transformation in Eq.10 is a rotation and the
boost a helicity boost, Eq.4, then $\Lambda (e_W)$ becomes a diagonal
matrix of phases, exactly as is the case for massive particles with
helicity boosts \cite{f}.

The transformation properties of photon wave functions are inherited from
those of the states.  If a state $|\phi>$ is written in terms of wave
functions and basis states,
\begin{eqnarray}
|\phi>&=&\sum\int \frac{d^3 k}{k_0} \phi(k,\alpha)|k,\alpha>,\
\end{eqnarray}
and the action of the Lorentz transformation in Eq.10 on states is
transferred to the wave function, one obtains
\begin{eqnarray}
(U_\Lambda\phi)(k,\alpha)&=&\sum\Lambda_{\alpha,\alpha^{'}}
(e_W(k,\Lambda^{-1}))\phi(\Lambda^{-1}k,\alpha^{'}).
\end{eqnarray}
Here $d ^3k/k_0$ is the Lorentz invariant measure with $k_0=|\vec
k|$.  To be more mathematically precise, the Hilbert space discussed in
the following paragraphs, and the transformation properties of the
Hilbert space elements under Lorentz transformations
given in Eq.13 could all be derived directly from induced representation
theory
\cite{g}.  The only somewhat unusual feature would be that the
representation of the Euclidian group is four dimensional, rather than
the more usual one dimensional representation.  Such a mathematical
background would be necessary to prove the irreducibility of the massless
representation, a subject that will not be pursued further in this paper.

To show that the representation defined by Eq.13 is unitary, it is
necessary to define an inner product on photon wave functions.  Since the
representation of the little group, Eqs.2,3, is not unitary, the usual
Hilbert space of square integrable functions will not lead to a unitary
representation for the Poincar\'{e} group.   But the representation of the
little group is in terms of Lorentz matrices which satisfy $\Lambda
g\Lambda^T=g$; this property can be used to define an inner product which
produces a unitary representation for the Poincar\'{e} group. Define a
photon inner product as
\begin{eqnarray}
(\phi,\psi):&=&-\sum\int \frac{d^3 k}{k_0}
\phi^{\ast}(k,\alpha)g_{\alpha,\alpha}\psi(k,\alpha);\
\end{eqnarray}
then the representation defined by eq.13 is unitary, that is
$||U_\Lambda\phi||^2=||U_a\phi||^2=||\phi||^2$.

But the inner product as defined in Eq.14 is not positive definite (for
example choose $\phi(k,0)=\phi(k)$, with the other components zero).  As is
well known \cite{c}, to get around this problem, the zero component  of
the wave function is chosen to equal the third component of the wave
function:
\begin{eqnarray}
\phi(k,0)&=&\phi(k,3),\\
\sum k^{st}_\alpha g_{\alpha,\alpha} \phi(k,\alpha)&=&0.\
\end{eqnarray}
Using Eq.13 it is easy to see that the condition, Eq.16, is Lorentz
invariant.  This allows one to define the photon Hilbert space $H_\gamma$ as
\begin{eqnarray}
H_\gamma&=&\{\phi(k,\alpha)|\phi(k,0)=\phi(k,3)\},\
\end{eqnarray}
with the inner product given in Eq.14, resulting in a positive
definite inner product, in which the zero and three components of the wave
function do not contribute to the inner product because of Eq.15. 
Moreover, quantities such as the expectation value of the energy,
$(\phi,P_0\phi)$ are also seen to be positive definite, with the value zero
occurring only when the one and two component parts of the wave function
are zero.

With this background it is now possible to introduce many photon states and
wave functions, living in the Fock space generated by sums of symmetrized
tensor products of $H_\gamma$, through photon creation and annihilation
operators, whose transformation properties are inherited from the one
particle photon properties:
\begin{eqnarray}
|k,\alpha>&=&c^\dagger(k,\alpha)|0>\\
c(k,\alpha)|0>&=&0, \alpha =0,1,2,3\\
 {[c(k,\alpha),c^\dagger(k^{'},\alpha^{'})]}&=&-g_{\alpha,\alpha^{'}} k_0
\delta^3(\vec k -\vec k^{'})\\
U_\Lambda c(k,\alpha)U_\Lambda^{-1}&=&\sum c(\Lambda
k,\alpha^{'})\Lambda_{\alpha,\alpha^{'}}(e_W)\\
U_ac(k,\alpha)U_a^{-1}&=&e^{-ik\cdot a} c(k,\alpha)\\
P^\gamma_\mu&=&-\sum\int \frac{d^3k}{k_0}k_\mu
c^\dagger(k,\alpha)g_{\alpha,\alpha} c(k,\alpha).\
\end{eqnarray}
The important difference between the usual Gupta-Bleuler analysis and
this paper is seen in Eq.21.  The creation and annihilation operators do
not transform as four-vectors under Lorentz transformations, but as
irreducible representations of the Poincar\'{e} group for massless
particles, in which the little group representation is a four dimensional
nonunitary representation of the two dimensional Euclidean group.

The wave function condition,Eq.15, now becomes the annihilation operator
condition
\begin{eqnarray}
\sum k^{st}_\alpha  c(k,\alpha)|\phi>=0,\
\end{eqnarray}
for all $k$ and for all $|\phi>$ in the Fock space. By applying a Lorentz
transformation to Eq.24 and using Eq.21, it is straightforward to show
that Eq.24 is a Lorentz invariant condition.  Moreover, it guarantees that
no timelike or longitudinal components will contribute to the inner
product.  That is, if $|\phi_n>$ is an n-photon state, Eq.24 guarantees
that the $\alpha_i =0$ components will cancel the $\alpha_i =3$ components in
the n-photon wavefunction.  Hence the inner product,
\begin{eqnarray}
<\phi_n|\phi_n>&=&(-1)^n \sum_{\alpha_i} \int\prod_{i} \frac{d^3
k_i}{k_{0i}} |\phi_n(k_1\alpha_1...k_n\alpha_n)|^2
g_{\alpha{_1},\alpha_1}...g_{\alpha{_n},\alpha_n}\nonumber\\
&=&\sum_{\alpha_i =1,2}\int\prod_{i} \frac{d^3 k_i}{k_{0i}}
|\phi_n(k_1\alpha_1...k_n\alpha_n)|^2,\
\end{eqnarray}
where all the $\alpha_i=0$ components have cancelled with the $\alpha_i=3$
components, so that the norm is always nonnegative.
\section{The Free Four-Vector Potential Operator}
In this section, as in all the papers in this series,  fields will be
defined as translates of the four-momentum operator from the space-time
point zero.   In this paper the four-momentum operator is
taken to be the photon four-momentum operator, defined in Eq.23, while in
the next paper \cite{h}, it will include matter and electromagnetic
four-momentum operators.  However, the four-vector potential operator at
the space-time point zero is always defined by
\begin{eqnarray}
A_\mu(0):&=&-\sum \int \frac{d^3k}{k_0} B_{\mu
\alpha}(k)g_{\alpha,\alpha}(c(k,\alpha)+c^\dagger(k,\alpha)).\
\end{eqnarray}
As a consequence of this definition and the transformation properties of
the photon creation and annihilation operators, Eq.21, the four-vector
potential operator will transform as a four-vector under Lorentz
transformations, $U_\Lambda A_\mu(0) U_\Lambda^{-1}=(\Lambda^{-1})_\mu^\nu
A_\nu(0)$.  Further the polarization vector, usually written as
$\epsilon_{\mu\alpha}(k)$ (see for example, Schweber, reference \cite{i},
page 249), is seen to be the boost matrix discussed after Eq.3.  Usually a
helicity boost is  chosen (see Eq.4) but it is clear that any other boost
will serve equally well, for example the boost defined in the appendix,
Eq.42.

The free four-vector potential operator at the space-time point $x$ is
defined to be
\begin{eqnarray}
A_\mu(x):&=&e^{iP^\gamma \cdot x} A_\mu(0) e^{-iP^\gamma \cdot x}\\
&=&-\sum \int \frac{d^3k}{k_0}  B_{\mu \alpha}(k)g_{\alpha,\alpha}
(e^{-ik\cdot x} c(k,\alpha)+e^{ik\cdot x} c^\dagger(k,\alpha)).
\end{eqnarray}
  From
this definition it follows that this operator is local, that is, the
commutator $[A_\mu(x),A_\nu(y)]$ is zero for $(x-y)^2$ spacelike. Eq.28
can also be used to relate the generalized Gupta-Bleuler formalism
developed in this paper with the usual Gupta-Bleuler formalism.  If the
boost in Eq.28 is chosen to be a helicity boost, Eq.5, then the
annihilation operator transforming as a four-vector under Lorentz
transformations is related to the annihilation operator transforming as a
one particle state under Lorentz transformations (Eq.21) by 
$c(k,\mu)=B_{\mu \alpha}(k)g_{\alpha,\alpha}c(k,\alpha).$

Finally, the positive frequency part of the four vector field
satisfies a (nonlocal) Lorentz gauge condition:
\begin{eqnarray} 
A_\mu ^+ (x):&=&-\sum \int \frac{d^3k}{k_0} B_{\mu \alpha}(k)
g_{\alpha,\alpha} e^{-ik\cdot x} c(k,\alpha)\\
\partial A_\mu^+ (x)/\partial x_\mu&=&i\sum \int \frac{d^3k}{k_0} k^\mu
B_{\mu
\alpha}(k) g_{\alpha,\alpha}e^{-ik\cdot x} c(k,\alpha)\nonumber\\
&=&i\sum \int \frac{d^3k}{k_0} e^{-ik\cdot x}k^{st}_\alpha
c(k,\alpha)\nonumber\\ &=&0,
\end{eqnarray}
where use has been made of the fact that an inverse boost on the four
vector $k$ results in the four vector $k^{st}$.

This section  concludes with a discussion of gauge invariance for the
creation and annihilation operators and the four-vector potentials. 
If $c(k,\alpha)$ is the annihilation operator defined in Eq.19 ff, then
the gauge transformed annihilation operator is defined to be
\begin{eqnarray}
c^{'}(k,\alpha):&=&c(k,\alpha)+k^{st}_\alpha f(k)I,
\end{eqnarray}
where $f(k)$ is a complex function of the four vector $k$ and $I$ is the
identity operator\footnote{It would also be possible to make $f(k)I$ an
operator, say $d(k)$, which acts on the full photon Fock space.  This is
equivalent to introducing an auxiliary massless scalar field.  Requiring
that
$c^{'} (k,\alpha)$ satisfy the same properties as $c(k,\alpha)$ then puts
requirements on $d(k)$ see reference \cite{j} and in particular page 75.
}. 
$c^{'}(k,\alpha)$ must satisfy the same conditions as
$c(k,\alpha)$, namely Lorentz covariance (Eq.21), boson commutation
relations (Eq.20), and the subsidiary condition (Eq. 24).  The subsidiary
condition for $c^{'}(k,\alpha)$ follows immediately from the definition,
since $k^{st}\cdot k^{st}=0$.  Also the commutation relations follow since
the term added to $c(k,\alpha)$ is a multiple of the identity operator.

The Lorentz condition can be written as
\begin{eqnarray}
U_\Lambda c^{'}(k,\alpha) U^{-1}_\Lambda &=&\sum c(\Lambda
k,\alpha^{'})\Lambda_{\alpha^{'} \alpha}(e_W)+k^{st}_\alpha f(k)I\nonumber\\
&=&\sum(c(\Lambda k,\alpha^{'})+k^{st}_{\alpha^{'}}
f(k)I)\Lambda_{\alpha^{'}
\alpha}(e_W)\nonumber\\
&=&\sum c^{'}(\Lambda k,\alpha^{'}) \Lambda_{\alpha^{'} \alpha}(e_W),
\end{eqnarray}
the desired result if $f(k)=f(\Lambda k)$, that is, $f(k)$ is a Lorentz
scalar.  Gauge transformations have the effect of adding or subtracting
equal amounts of  timelike and longitudinal components and thus do
not change the norm of many-photon wave functions.

Finally, the positive frequency part of the gauge transformed four vector
field is also conserved:
\begin{eqnarray}
A_\mu ^{{'}+}(x):&=&\sum \int \frac{d^3 k}{k_0} B_{\mu\alpha}(k)
e^{-ik\cdot x}g_{\alpha,\alpha} c^{'}(k,\alpha)\\
&=&A^+_\mu(x)+\sum\int \frac{d^3 k}{k_0} k_\mu e^{-ik\cdot x}
f(k)I;\nonumber\\
\partial A^{{'}+}_\mu(x)/\partial x_\mu&=&\partial A^{+}_\mu(x)/\partial
x_\mu-i\sum\int \frac{d^3 k}{k_0} k\cdot k e^{-ik\cdot x} f(k)
I\nonumber\\ &=&0.
\end{eqnarray}
\section{Conclusion}

In order to be able to write electromagnetic interactions as vertex
interactions, it is necessary to generalize the Gupta-Bleuler formalism
so that photon creation and annihilation operators transform as single
particle states under Lorentz transformations.  As first pointed out by
Wigner, the little group for massless particles is the two dimensional
Euclidian group E(2);  but the choice of representation for E(2) is not
given a priori.  Wigner
\cite{e}(and later others, including Weinberg, reference \cite{d}) choose
the degenerate one dimensional representation of E(2), in which the
action of the translations in E(2) are trivial.  While it is possible to
obtain the usual photon states and wavefunctions with such a
representation of E(2), troubles arise not only with gauge invariance,
but also with providing the natural link between fields and photon
creation and annihilation operators, of the sort available for massive
particles with spin (see for example, reference \cite{d}, page 233).

In this paper the massless little group representation is chosen to be the
four dimensional nonunitary representation of E(2), obtained by
restricting elements of the Lorentz group to E(2); the form of these E(2)
elements, as Lorentz transformations, is given in Eqs.2,3.   Such a
representation must be nonunitary, since it is a finite dimensional
representation of a noncompact group.  Nevertheless the representation of
the full Poincar\'{e} group is unitary.  Such a result makes use of the
fact that E(2) matrices are Lorentz matrices;  by suitably  modifying the
inner product, the resulting Poincar\'{e} representation is unitary and
the inner product agrees with that given by the Gupta-Bleuler formalism.

The inner product that makes the representation of the Poincar\'{e} group
unitary is not positive definite.  So-as with the Gupta-Bleuler formalism-
the photon Hilbert space is defined as the subspace of wavefunctions for
which the timelike and longitudinal components are equal.  Such a
subspace is a Poincar\'{e} invariant subspace.  Many-photon states and
wavefunctions can then be defined in terms of photon creation and
annihilation operators.  These creation and annihilation operators do not
however transform as four vectors, as is usually the case (see reference
\cite{i},page 243);  rather under Lorentz transformations they transform
as Euclidian analogues of Wigner rotations (see Eq.21), which is the
natural generalization of the transformation properties for massive
particles with spin.  Because of these transformation properties, the
proof of the operator condition that longitudinal and time-like
components cancel is Lorentz invariant is particularly simple.

The main result of this paper concerns the link between the
four-vector potential operator and photon creation and annihilation
operators.  For massive particles with spin this link is always given by
Lorentz group representations of boosts, coset representatives of SO(1,3)
with respect to the rotation group.  For example the usual spinor
functions for spin 1/2 fermions are boosts, usually canonical spin
boosts.  But there are many boost possibilities, such as helicity or
front form boosts \cite{k}.  Similarly for massless particles boosts are
coset representatives of SO(1,3) with respect to E(2), and provide the
link between the four-vector potential and photon creation and
annihilation operators.  That is, polarization vectors are boost
representatives, coset choices of SO(1,3) with respect to E(2).  The
usual polarization choice is the helicity boost, given in Eq.5.  But just
as there are many different boost choices for massive particles, all
connected by Melosh rotations \cite{k}, \cite{l}, so too there are many
boost choices for massless particles, all connected by the analogue of
Melosh rotations, namely E(2) transformations.  An example of a
non-helicity polarization vector, a front form boost for massless
particles in given in the appendix, Eq.42.

Gauge transformed photon creation and annihilation operators with the
correct Lorentz (Eq.21) and subsidiary (Eq.24) conditions affect only the
time-like and longitudinal polarizations, leaving the transverse parts
unchanged.  Using the connection between four-vector potentials and
creation and annihilation operators, the usual gauge transformations for
the positive frequency part of the four-vector potentials are obtained,
as well  as the fact that under such a gauge transformation the Lorentz
gauge condition remains invariant.  Since the four-vector potential
operator transforms as a four-vector under Lorentz transformations, it
can be coupled to the current operators defined in the previous paper, to
form the electromagnetic vertex for particles of any spin, which is the
starting point for constructing the electromagnetic mass operator.

To conclude it should be pointed out that the procedures
applied here to photons can equally well be applied to massless spin two
(or for that matter to arbitrary spin) particles, namely gravitons.  The
construction of the relevant nonunitary representations of the Euclidian
group, as well as the construction of different polarization tensors,
will be discussed in another paper.
\appendix
\section{Appendix: SL(2,C) and Massless Particles}  

In section 2 all operations were carried out using the Lorentz group
SO(1,3).  In this appendix the fact that SL(2,C) is the covering group of
the Lorentz group is used to derive some results in a more transparent
way.  Under a Lorentz transformation $\Lambda$, a four vector $k$ goes to
$k^{'}=\Lambda k$.  Such a transformation is carried out in
SL(2,C) by replacing the four-vector $k$ by the hermitian matrix $H(k)$, with
\begin{eqnarray}
H(k^{'})&=&A H(k) A^\dagger\\
H(k)&=&\left[\begin{array}{cc}k_+&k{_\perp ^\ast}\\
k_\perp&k_-\\
\end{array}\right].\
\end{eqnarray}
$A$ is an element of SL(2,C) and $k_\pm=k_0\pm k_z, k_\perp=k_x +ik_y$. 

 The
standard vector $k^{st}$ introduced in Eq.3 becomes $H(k^{st})$ and the
little group is the subgroup of SL(2,C) that leaves $H(k^{st})$ invariant:
\begin{eqnarray}
E(2)&=&\{A\in SL(2,C)|H(k^{st})=AH(k^{st})A^\dagger\}\\
&=&\{e_2(\phi,a)\}\nonumber\\
&=&\{\left[\begin{array}{cc}e^{i\phi/2}&ae^{i\phi/2}\\
0&e^{-i\phi/2}
\end{array}\right]\},\
\end{eqnarray}
where now $a=a_x+ia_y$ gives the two translations.
Written in this way it is straightforward to show that the elements in 
Eq.38 combine as E(2) elements.

Any element of SL(2,C) can be decomposed into boosts (coset
representatives) with respect to E(2).  A natural choice is a front form
boost,
\begin {eqnarray}
B_F(k):&=&\left[\begin{array}{cc}\sqrt{k_+/2}&0\\
k_\perp/\sqrt{2k_+}&\sqrt{2/k_+}\\
\end{array}\right]\\
H(k)&=&B_F(k) H(k^{st})B_F^\dagger(k),\
\end{eqnarray}
for then
\begin{eqnarray}
A&=&B_F(k)e_2(\phi, a),\
\end{eqnarray}
and the parameters of A are readily expressed in terms of $k$ and the
Euclidian parameters $\phi$ and $a$.

The correspondence between elements of SL(2,C) and SO(1,3) \cite{m} then
gives the front form boost, Eq.39, as an SO(1,3) element, suitable for a
polarization vector:
\begin{eqnarray}
B_F(k)&=&\left[\begin{array}{cccc}k_0/2+1/k_+&k_x/k_+&k_y/k_+&k_0/2
-1/k_+\\k_x/2&1&0&k_x/2\\
k_y/2&0&1&k_y/2\\k_z/2-1/k_+&-k_x/k_+&-k_y/k_+&k_z/2+1/k_+\\
\end{array}\right].\
\end{eqnarray}
Since an arbitrary Lorentz group element can now be decomposed in (at
least) two ways, namely $\Lambda=B_H(k)e^{'}_2=B_F(k)e_2$, it follows that
$B_F(k)=B_H(k)e_2^{'}e_2^{-1}$, that is, all boosts are related by $e_2$
elements, which is the massless analogue of Melosh rotations \cite{l}.


\begin{thebibliography}{99}
\bibitem{a}W. H. Klink, "Constructing Point Form Mass Operators from
Vertex Interactions", submitted.
\bibitem{b}W. H. Klink, "Local Current Operators for Arbitrary Spin
Particles", submitted.
\bibitem{c}K. Bleuler,  Helv.Phys.Acta, {\bf23} (1950) 567, S. N. Gupta,
Proc. Phys. Soc. (London), {\bf A63} (1950) 681;  see also S. S.
Schweber, \textsl{An Introduction to Relativistic Quantum Field Theory}
 (Harper and
Row, New York, USA, 1962), chapter 9.
\bibitem{d}S.  Weinberg, \textsl{ The Quantum Theory of Fields, Volume I}
(Cambridge University Press, Cambridge, England, 1995).
\bibitem{e}E. P. Wigner, Ann. Math. {\bf{40}} (1939) 149.
\bibitem{f}W. H. Klink, Ann. Phys. (N.Y.) {\bf213} (1992) 31.
\bibitem{g}R. Warren, W. H. Klink, J. Math. Phys. {\bf{11}} (1970) 1155; 
G. Mackey, \textsl {The Theory of Induced Representations} (University of
Chicago Press, Chicago, USA, 1955).
\bibitem{h}W. H. Klink, "Point Form Electrodynamics and the Construction
of Conserved Current Operators", submitted.
\bibitem{i}See for example, Schweber, reference \cite{c}, page 243.
\bibitem{j}N. Nakanishi, I. Ojima, \textsl {Covariant Operator Formalism
of Gauge Theories and Quantum Gravity} (World Scientific Lecture Notes in
Physics, Volume 27, Singapore, 1990).
\bibitem{k}W. H. Klink, Phys.Rev.C{\bf58} (1998) 3617.
\bibitem{l}H. D. Melosh, Phys.Rev.D{\bf9} (1974) 1095;  also reference
11, page 3625.
\bibitem{m}M.A. Naimark, \textsl{Linear Representations of the Lorentz
Group}, (Pergamon Press, Oxford, England, 1964), page 122.
\end{thebibliography}
\end{document}